\documentclass[twocolumn,twocolappendix,times,tighten,10pt]{aastex631}
\usepackage{amsmath}
\usepackage{amssymb}
\usepackage{bm}

\newcommand{\Machnum}{\mathcal{M}}

\begin{document}

\title{Supersonic Turbulence in Primordial Halos: A Comparison With and Without The Stream Velocity}

\correspondingauthor{Avi Chen}

\author[0000-0002-8859-7790]{Avi Chen}
\affiliation{Department of Physics and Astronomy, Rutgers, The State University of New Jersey, 136 Frelinghuysen Rd, Piscataway, NJ 08854, USA \\}
\email{avi.chen@rutgers.edu}

\author[0000-0002-4227-7919]{William Lake}
\affil{Department of Physics and Astronomy, UCLA, Los Angeles, CA 90095}
\affil{Mani L. Bhaumik Institute for Theoretical Physics, Department of Physics and Astronomy, UCLA, Los Angeles, CA 90095, USA\\}

\author[0000-0003-2369-2911]{Claire E. Williams}
\affil{Department of Physics and Astronomy, UCLA, Los Angeles, CA 90095}
\affil{Mani L. Bhaumik Institute for Theoretical Physics, Department of Physics and Astronomy, UCLA, Los Angeles, CA 90095, USA\\}

\author[0000-0001-5817-5944]{Blakesley Burkhart}
\affiliation{Department of Physics and Astronomy, Rutgers, The State University of New Jersey, 136 Frelinghuysen Rd, Piscataway, NJ 08854, USA \\}
\affiliation{Center for Computational Astrophysics, Flatiron Institute, 162 Fifth Avenue, New York, NY 10010, USA \\}

\author[0000-0002-9802-9279]{Smadar Naoz}
\affil{Department of Physics and Astronomy, UCLA, Los Angeles, CA 90095}
\affil{Mani L. Bhaumik Institute for Theoretical Physics, Department of Physics and Astronomy, UCLA, Los Angeles, CA 90095, USA\\}

\author[0000-0002-0311-2206]{Shyam H. Menon}
\affiliation{Department of Physics and Astronomy, Rutgers, The State University of New Jersey, 136 Frelinghuysen Rd, Piscataway, NJ 08854, USA \\}
\affiliation{Center for Computational Astrophysics, Flatiron Institute, 162 Fifth Avenue, New York, NY 10010, USA \\}

\author[0000-0003-3816-7028]{Federico Marinacci}
\affiliation{Department of Physics \& Astronomy ``Augusto Righi", University of Bologna, via Gobetti 93/2, 40129 Bologna, Italy\\}
\affiliation{INAF, Astrophysics and Space Science Observatory Bologna, Via P. Gobetti 93/3, I-40129 Bologna, Italy\\}

\author[0000-0001-8593-7692]{Mark Vogelsberger}
\affil{Department of Physics and Kavli Institute for Astrophysics and Space Research, Massachusetts Institute of Technology, Cambridge, MA 02139, USA\\}

\author[0000-0001-7925-238X]{Naoki Yoshida}
\affiliation{Department of Physics, The University of Tokyo, 7-3-1 Hongo, Bunkyo, Tokyo 113-0033, Japan}
\affiliation{Kavli Institute for the Physics and Mathematics of the Universe (WPI), UT Institute for Advanced Study, The University of Tokyo, Kashiwa, Chiba 277-8583, Japan}
\affiliation{Research Center for the Early Universe, School of Science, The University of Tokyo, 7-3-1 Hongo, Bunkyo, Tokyo 113-0033, Japan}

\begin{abstract}

Turbulence plays a critical role in regulating star formation in molecular clouds and is also observed in simulations of primordial halos that host Population III (Pop III) stars. The relative velocity between baryons and dark matter at the time of recombination is thought to be a source of turbulence in the early universe. In this paper, we study how this stream velocity affects the turbulence inside primordial halos using high-resolution cosmological simulations across the redshift range of $z = 30$ to $z = 20$. We find that at a fixed redshift, the stream velocity enhances turbulence in low-mass halos ($M \lesssim 10^6 \ \mathrm{M_\odot}$) and suppresses it for high-mass halos ($M \gtrsim 10^6 \ \mathrm{M_\odot}$). The enhancement in low-mass halos likely arises from residual kinetic energy introduced by the stream velocity, while the suppression in high-mass halos likely arises from a reduction in inflowing accretion-driven turbulence. This mass-dependent modulation of turbulence suggests that the initial conditions inside primordial halos are altered in the presence of the stream velocity, potentially influencing their fragmentation and the resulting star formation.

\end{abstract}


\section{Introduction}\label{sec:intro}

For the first $\sim$ 380,000 years after the Big Bang the Universe was mostly a homogeneous and isotropic plasma containing protons, neutrons, electrons, and photons. The baryons were coupled to the photons via their mutual coupling to the electrons. Slight baryonic overdensities, originally seeded by quantum fluctuations, were therefore prevented from growing gravitationally due to the radiation pressure of the photons. This created a standing wave of matter known as Baryonic Acoustic Oscillations (BAO). Since dark matter (DM) does not interact with electromagnetic radiation it did not experience these oscillations and accelerated gravitationally while coalescing into potential wells. This caused a significant relative velocity between the DM and the baryons. At the time of recombination, $z \sim 1020$, the rate of Thomson scattering dropped dramatically, and the photons kinetically decoupled from the baryons (though they were still somewhat coupled through scattering with the residual uncombined electrons). As a result, the sound speed of the baryons dropped from highly relativistic \( \sim c/\sqrt{3} \) to thermal speeds of an ideal gas \( \sim 6 \, \text{km/s} \). The DM, not being coupled to the radiation, did not experience this negative acceleration. The relative velocity, therefore, became supersonic with (following a Gaussian distribution) rms velocities $\sigma_{\rm vbc}$ \( \sim 30 \, \text{km/s} \) and a Mach number of $\mathcal{M}$ $\sim$ 5 \citep[e.g.,][]{Tes+10a,Tes+10b}. 

Using moving-background perturbation theory, \cite{Tes+10a} showed the necessity of including the term representing the relative velocity between the baryons and the DM at the time of recombination despite it being nominally second-order. This is because at the scale $k_{vbc} \equiv aH/v_{bc}$, the expansion parameter of this term is of order unity \citep{Tes+10b}. Being ultimately derived from similar physics concepts as the BAO, the relative velocity is coherent on the same scales as the Silk damping of a few Mpc and can therefore be numerically modeled as a uniform velocity boost below that scale. The relative velocity, therefore, is often simply called the ``stream velocity'' or ``streaming'' and decays, like the peculiar velocity of a free particle, as $a(t)^{-1}$. 

The stream velocity alters the formation and observational signatures of early-universe objects at high redshifts. The spatial offset between baryons and DM generated by the stream velocity acts as an effective anisotropic pressure on the gas \citep[e.g.,][]{McOL12}. This has a cascading effect that has been well studied both analytically and numerically (see the review by \citealt{Fialkov14}). The minimum halo mass required for cooling and the Jeans mass is increased, delaying the formation of Population III stars (Pop III) \citep[e.g.,][]{Fialkov+11, Stacy+10,Greif+11,Maio+11,Naoz+11a,Naoz+12,Popa+15,Schauer+17a,Hegde_23}. Inside minihalos the gas fraction, density, and cold gas decreases, while the spin and rotational support of the halo increases \citep[e.g.,][]{Naoz+12, McOL12, Richardson+13, Schauer+19, Chiou+18, Williams+23}. Streaming delays DM halo collapse, suppresses the cumulative halo mass function, generates gas empty halos, and decreases the number of low-mass luminous galaxies \citep[e.g.,][]{ Naoz+11a, BD, Asaba+16}. Supermassive black hole formation can be enhanced, potential globular cluster progenitors can form, and some may be detectable by JWST \citep[e.g.,][]{Latif+14, Tanaka+14, Hirano+17, Lake+21, Nakazato+22, Lake+23a, Lake+23b, Lake_25}. The 21 cm signal before the formation of the first stars is enhanced/suppressed at different scales, and large-scale fluctuations are altered during reionization \citep[e.g.,][]{Ali14, McOL12, Visbal+12, Munoz+19, Cain+20, Park_21}. The BAO peak can be shifted in both Fourier space, configuration space, and when using the 21 cm signal as a probe \citep[e.g.,][]{Yoo+11, Dalal+10, Slepian15, Long+22}. The faint end of the UV luminosity function, as well as the star formation rate in dwarf galaxies, can also be enhanced at high redshift \citep[e.g.,][]{Williams_2024, Lake+24a}.

The stream velocity is also expected to inject turbulence into the gas of primordial halos that host Pop III stars \citep{Glover23}. However, aside from \citet{Greif+11}, which showed that the stream velocity enhanced turbulence in the first collapsing minihalos in their simulations, no study has systematically examined how the stream velocity influences turbulence across a broad population of halos at the onset of Pop III star formation, nor explained what determines the resulting turbulence levels. One other exception is \citet{Latif_14}, who found no significant impact of the stream velocity on turbulent energy, density, or radial velocities in halos at $z = 10$. However, this null result is not surprising, as the stream velocity has largely decayed by that redshift and is no longer dynamically important. 

Turbulence is important and plays a key role in many astrophysical processes, including cosmic ray transport \citep[e.g.,][]{Lazarian_06, Lazarian_23}, regulation of the ISM \citep[e.g.,][]{Falceta_14, Ho_23, Yuen_24}, and star formation in local molecular clouds \citep[e.g.,][]{Krumholz_05, Burkhart18}. Turbulence also plays a critical role in the formation of Pop III stars. Supersonic turbulence in primordial halos is generated during the virialization process and gravitational infall \citep[e.g.,][]{Wise07, Greif08, Chen_25, Ho_25}. This turbulence can lead to fragmentation of the cloud itself \citep[e.g.,][]{Sugimura_2023} or the protostellar disk, with the extent of fragmentation being influenced by both the magnitude of the turbulent energy and its mode $-$ solenoidal versus compressive $-$ and the rotational energy in the cloud \citep{Clark11b, Riaz18, Riaz23, Wollenberg2020}. Turbulence can also amplify the magnetic fields by stretching and folding the field lines, which was shown to have a significant impact of the IMF of the Pop III stars \citep[e.g.,][]{Sharda_20, 2021MNRAS.503.2014S, Sharda_24}, and is the main source of angular momentum transport during the initial collapse \citep[e.g.,][]{Yoshida+06, Greif_12}.

Turbulence in these primordial minihalos also resembles that observed in local molecular clouds and idealized simulations. \citet{Prieto2011} found that the velocity dispersion within these minihalos on scales of 100 pc is comparable to that in local molecular clouds. Additionally, the slope of the second-order structure function is marginally steeper than that observed in randomly driven supersonic turbulence \citep{Prieto2011}. The probability density function (PDF) of density in these halos exhibits a log-normal distribution, with parameters akin to those in supersonic turbulence simulations \citep{Safranek-Shrader}. In the local universe the density PDF is crucial for characterizing the star formation rate \citep[e.g.,][]{Burkhart+18, Appel_23}.

In this paper we study how the stream velocity modulates this turbulence by looking at its impact over a range of halos with different masses during the epoch of Pop III star formation. The paper is structured as follows. In $\S$~\ref{sec:numerical}, we describe the numerical set-up of our simulations, including the implementation of the stream velocity. In $\S$~\ref{sec:results}, we show how the stream velocity changes the turbulent velocities in primordial halos, analyze its dependence on halo mass and redshift, and explain its physical origins. In $\S$~\ref{sec:discussion}, we interpret our findings in the context of previous work, explore their implications for Pop III star formation and accretion physics, and discuss possible observational consequences. In $\S$\ref{sec:conclusion}, we summarize our key results.

We assume a $\Lambda$CDM cosmology, with $\Omega_{\rm \Lambda} = 0.73$, $\Omega_{\rm m} = 0.27$, $\Omega_{\rm b} = 0.044$, $\sigma_8  = 1.7$, and $h = 0.71$.
 
\section{Numerical Methods}
\label{sec:numerical}
\subsection{Initial Conditions}
\label{sec:IC}

In order to correctly set the initial conditions (ICs) of our simulation the perturbations of the Cosmic Microwave Background (CMB) need to be self-consistently evolved with the streaming effect. Without this correction, the resulting simulation will overestimate the gas density amplitude in the power spectrum of select wavenumbers and can underestimate the distance the gas is shifted \citep{Naoz+14, 2020ApJ...900...30P}.

To ensure that the streaming effect is realized self-consistently we generate separate transfer functions for DM and baryons by modifying the transfer functions from the CMBFAST code \citep{1996ApJ...469..437S}. We add the streaming effect given in equation (5) of \cite{Tes+10b} and add the scale-dependent temperature fluctuations $\delta_{\rm T}$ discussed in \cite{NB05}. The resulting coupled perturbation equations can be written as \citep{Naoz+12}:

\begin{equation}
\begin{aligned}
&\ddot{\delta}_{\rm c} + 2H\dot{\delta}_{\rm c} 
- f_{\rm c} \frac{2i}{a} \mathbf{v}_{\rm bc} \cdot \mathbf{k} \dot{\delta}_{\rm c} \\
&= \frac{3}{2} H_0^2 \frac{\Omega_{\rm m}}{a^3} (f_{\rm b} \delta_{\rm b} + f_{\rm c} \delta_{\rm c}) 
+ \left(\frac{\mathbf{v}_{\rm bc} \cdot \mathbf{k}}{a}\right)^2 \delta_{\rm c} \ ,
\end{aligned}
\label{eq:1}
\end{equation}

\begin{equation}
\begin{aligned}
\ddot{\delta}_{\rm b} + 2H\dot{\delta}_{\rm b} 
&= \frac{3}{2} H_0^2 \frac{\Omega_{\rm m}}{a^3} (f_{\rm b} \delta_{\rm b} + f_{\rm c} \delta_{\rm c}) \\
&\quad - \frac{k^2}{a^2} \frac{k_{\rm B} \bar{T}}{\mu} (\delta_{\rm b} + \delta_{\rm T}) \ ,
\end{aligned}
\label{eq:2}
\end{equation}
and the evolution of $\delta_{\rm T}$ can be related to the evolution of the density by:

\begin{equation}
\dot{\delta}_{\rm T} 
=  \frac{2}{3}\dot{\delta}_b + \frac{x_e}{t_\gamma a^4} \left[ \frac{\bar{T}_\gamma}{\bar{T}}( \delta_{T_\gamma} +  \delta_\gamma - \delta_{\rm T}) - \delta_\gamma \right] \ .
\label{eq:3}
\end{equation}

In these equations, $f_b$ and $f_{c}$ represent the cosmic baryon and cold DM fractions, while the variables $\delta_b$, $\delta_{c}$, $\delta_\gamma$, $\delta_{T_\gamma}$, and $\delta_{\rm T}$ denote the fluctuations in the baryon density, cold DM density, photon density, photon temperature, and baryon temperature, respectively. The mean photon and baryon temperatures are denoted by $\bar{T}_\gamma$ and $\bar{T}$. The quantity $x_e$ represents the electron fraction out of the total number density of gas particles, with its time dependence described in \citet{NN13}. The Thomson scattering coupling time is defined as: 
\begin{equation}
t_\gamma^{-1} \equiv \frac{8}{3} \bar{\rho}_\gamma^0 \frac{\sigma_T c}{m_e} = 8.55 \times 10^{-13} \, \mathrm{yr}^{-1}\ ,
\end{equation}
and it governs the thermal coupling between baryons and the CMB. $\mathbf{v}_{\rm bc}$ is the average relative velocity between the baryons and DM in a given patch of space, and $\mathbf{k}$ is the comoving wavenumber vector. $H$ is the Hubble parameter, $H_0$ is its present-day value, $\Omega_{\rm m}$ is the present-day matter density parameter, $a$ is the scale factor, $\mu$ is the mean molecular weight, and $k_{\rm B}$ is the Boltzmann constant.

\subsection{Cosmological Simulations}
\label{subsection:cosmological}

For our simulations, we use the quasi-Lagrangian code {\tt AREPO} \citep{2020ApJS..248...32W} that solves the equations of collisional and collisionless particles on a uniformly expanding, flat Friedmann-Lemaitre-Robertson-Walker spacetime. The mesh in {\tt AREPO} moves with the flow and is constructed as a Voronoi tessellation of the simulation volume starting from a set of mesh generating points. 

The gas in the simulation cools using the primordial chemistry and cooling library {\tt Grackle} \citep{2017MNRAS.466.2217S, Chiaki+19}. {\tt Grackle} tracks 49 nonequilibrium chemical reactions and their corresponding radiative cooling for the following 15 species: e$^-$, H, H$^+$, He, He$^+$, He$^{++}$, H$^-$, H$_2$, H$_2^+$, D, D$^+$, HD, HeH$^+$, D$^-$, and HD$^+$, with the most important lines being H$_2$ and HD cooling. The rate of radiative cooling of H$_2$ is computed for line transition between 20 rotational and 3 vibrational levels, and for HD for 3 vibrational levels.  

Our 2 Mpc$^3$ box contains 512$^3$ Voronoi gas cells with an initial mass of $m_{\rm gas}= 360\, \mathrm{M_\odot}$ and 512$^3$ DM particles with initial mass of $m_{\rm DM}= 1.9 \times 10^3\, \mathrm{M_\odot}$. The simulation starts at $z = 200$ and terminates at $z = 20$. We run the simulation twice: one run includes the stream velocity and one does not (with different ICs for each). Since the size of our box is less than the coherence scale of the stream velocity, we can implement the stream velocity by giving each gas cell a uniform velocity boost in the $\hat{x}$-direction. In this work we mainly study a rare patch of the universe that has a stream velocity of $v_{\rm bc} = 2\sigma_{\rm vbc}$, where $\sigma_{\rm vbc}$ is the (scale independent) rms fluctuation of the stream velocity. The value of $v_{\rm bc} = 2\sigma_{\rm vbc}$ allows for the full effect of streaming to be realized including the formation of gas structure outside the virial radius of DM halos \citep[also known as Supersonically Induced Gas Objects or SIGOs; e.g.,][]{Lake+21} and is similar to the estimated local value of $v_{\rm bc}=1.75^{+0.13}_{-0.28}\sigma_{\rm vbc}$, based on simulating Milky-Way-like galaxies with varying streaming velocities \citep{Uysal+22}. We set $\sigma_8  = 1.7$ to accelerate structure formation in our relatively small box. This has the effect of increasing the normalization of the halos abundances \citep{Naoz+12} and the time objects form by a factor of $\sqrt{2}$ \citep{Park+20} but does not otherwise affect the physics \citep[e.g.,][]{Yoshida+07,Stacy_10, Stacy_12, Susa_14, Greif+11}. At initialization we use a constant temperature of 422 K derived from linear theory, a softening length of 30 $h^{-1}$ pc, and set $\gamma$ = 5/3. The simulations do not include magnetic fields or star formation with its accompanying feedback. 

We use a standard friends-of-friends (FoF) algorithm to find halos and their associated gas cells \citep{Springel_01}, although this method has its limitations \citep[see][]{Wil+25}. We first run the FoF on the DM particles using a linking length of 20\% of the mean particle separation to identify DM halos. We then attach the gas cells within the halo in a secondary linking stage \citep{Popa+15}. We also compute the virial radius (assuming sphericity even though halos are triaxial \citep[e.g.,][]{Vogelsberger+20}) of the halo as the radius enclosing a mean density of 200 times the critical density of the Universe. In our subsequent analysis, we only consider DM halos that contain at least 300 DM particles and 100 gas cells.  

\section{Results}
\label{sec:results}

\subsection{Turbulence Inside Primordial Halos}
\label{subsec:turb}

To quantify turbulence within halos, we subtract the bulk, radial, and azimuthal rotational components from the velocities of the gas. The remaining random component of the velocity defines the turbulence:
\begin{equation}
v_{\mathrm{turb}} = \left( \frac{ \sum_i m_i \left| \bm{v}_i - \bm{v}_{\mathrm{rot},i} - \bm{v}_{\mathrm{rad},i} \right|^2 }{ \sum_i m_i } \right)^{1/2} \ ,
\label{eq:turbulent_velocity}
\end{equation}
 where the index $i$ runs over all gas cells, $m_i$ is the gas mass, ${\bm v}_i$ is the velocity relative to the center-of-mass velocity of the halo,  ${\bm v}_{\text{rad}, i} = ({\bm v}_i \cdot \hat{{\bm r}}_i) \hat{{\bm r}}_i$ is the radial component of the velocity and ${\bm v}_{\text{rot}, i} = {\bm r}_i \times {\bm \Omega}$ is the azimuthal rotational component of the velocity. ${\bm \Omega} = {\bm L} / I$ is the angular velocity of the gas and ${\bm L}$ is the total angular momentum of the gas given by $\sum_i m_i ({\bm r}_i \times {\bm v}_i)$, where and $I = \sum_i m_i |{\bm r}_i \times \hat{{\bm L}}|^2$ is the moment of inertia along the axis defined by the direction of $\bm{L}$ (see \citealt{2012MNRAS.426.1159S}). For each halo we also compute a root-mean-square (RMS), radial, and rotational velocities, given respectively as:
\begin{equation}
v_{\text{rms}} = \left(\frac{\sum_{i} m_i |{\bm v}_i|^2}{\sum_{i} m_i}\right)^{1/2} \ ,
\end{equation}
\begin{equation}
v_{\text{rot}} = \left(\frac{\sum_i m_i |{\bm v}_{\text{rot}, i}|^2}{\sum_i m_i}\right)^{1/2} \ ,
\label{eq:rotational_velocity}
\end{equation}
and
\begin{equation}
v_{\text{rad}} = \left(\frac{\sum_i m_i |{\bm v}_{\text{rad}, i}|^2}{\sum_i m_i}\right)^{1/2} \ .
\label{eq:radial_velocity}
\end{equation}

\begin{figure}[htbp]
    \centering
    \includegraphics[width=\columnwidth]{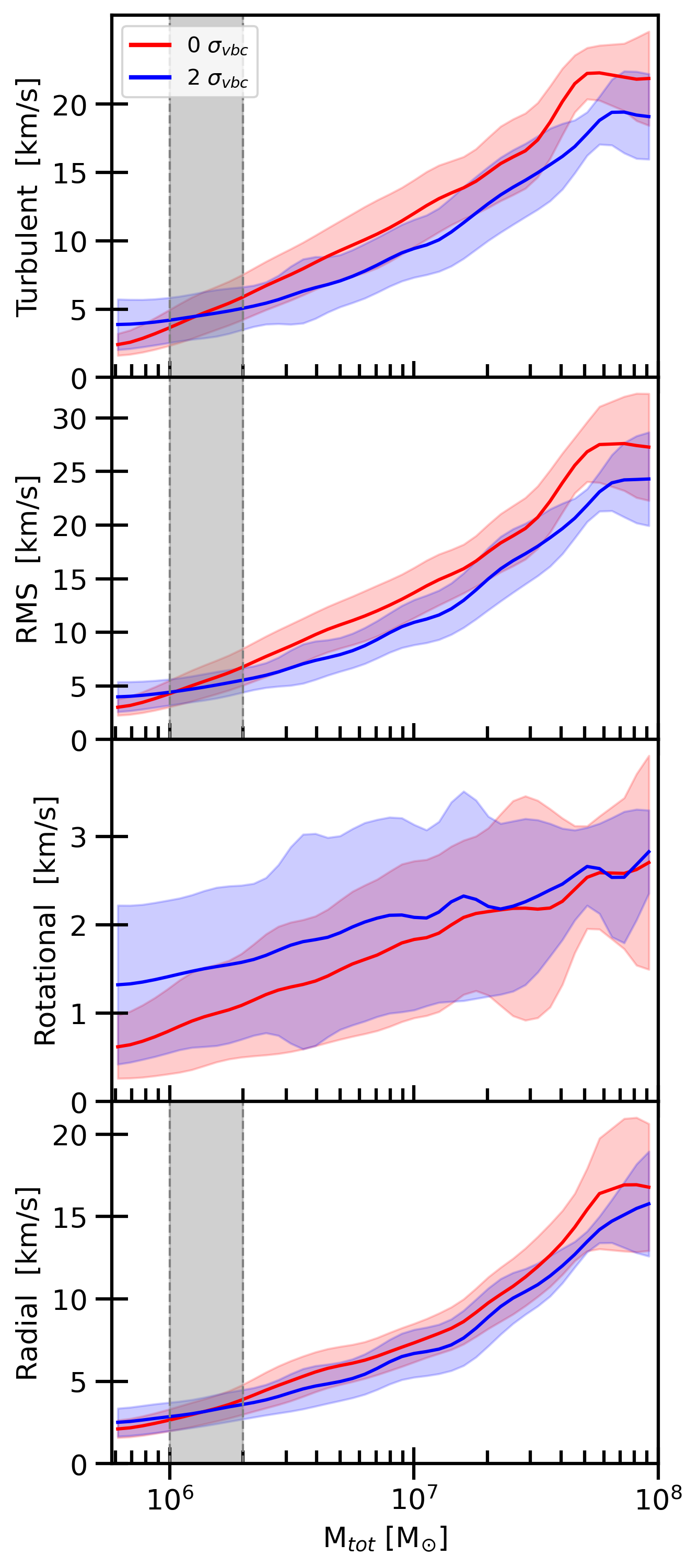}
    \caption{Comparison of kinematic properties of gas within halos in streaming ($v_{\rm bc} =2\sigma_{\rm vbc}$, blue) and no-streaming ($v_{\rm bc} = 0\sigma_{\rm vbc}$, red), simulations at $z = 20$. Panels from top to bottom show turbulent velocity, RMS velocity , radial velocity, and rotational velocity, each plotted against total halo mass (gas+DM). Shaded color bands indicate the $1\sigma$ scatter around the mean values. Streaming enhances turbulent and RMS velocities in low-mass halos ($M \lesssim 10^6 \ \mathrm{M_\odot}$) but suppresses them in high-mass halos ($M \gtrsim 10^6 \ \mathrm{M_\odot}$), with a clear turnover around the characteristic mass scale marked by the gray shaded band. Rotational velocities are amplified across all halo masses in the presence of streaming, while radial velocities are higher in high-mass halos without streaming.}
    \label{fig:vel_comp}
\end{figure}

\begin{figure}[ht!]
    \centering
    \includegraphics[width=\columnwidth]{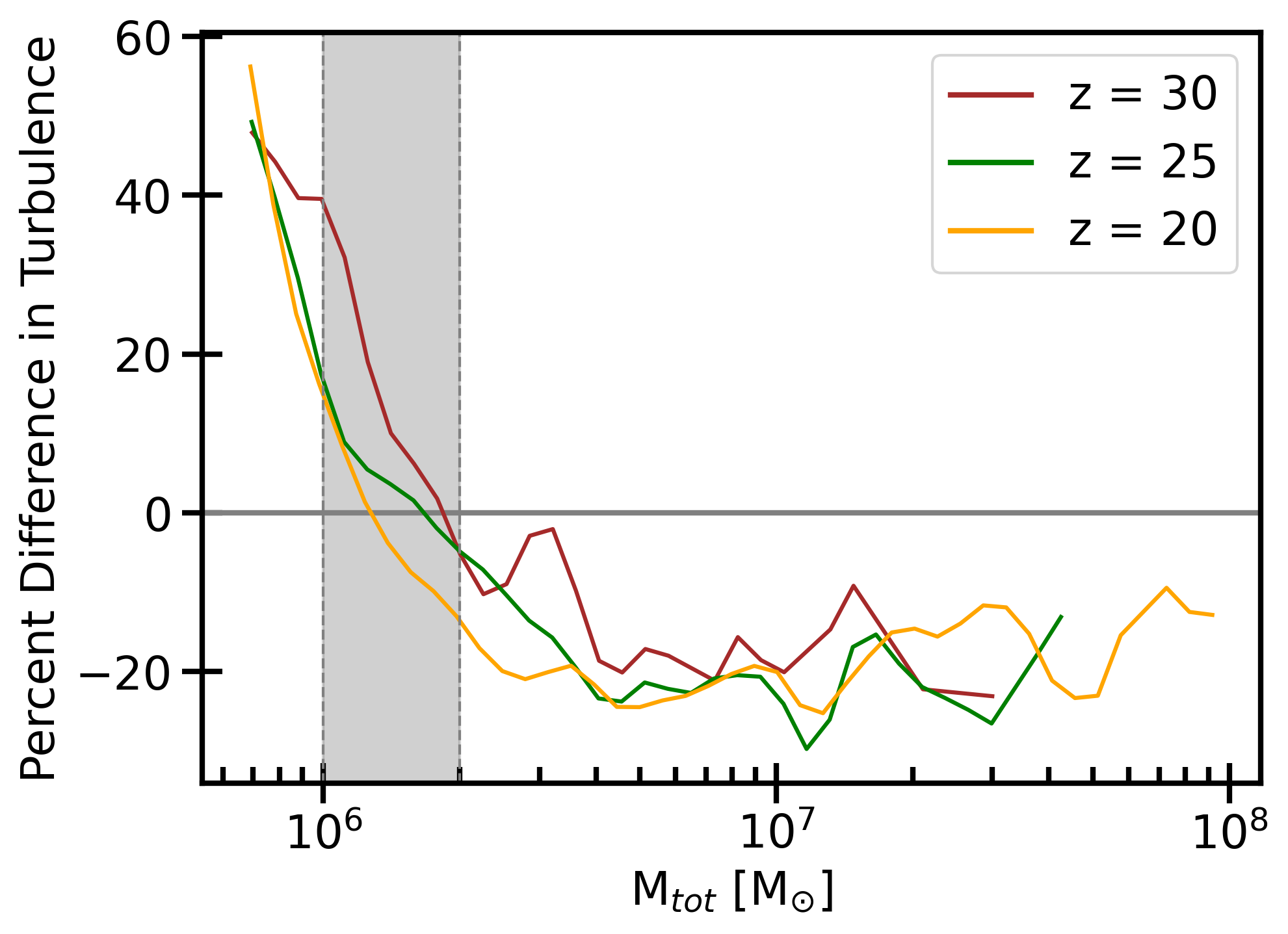}
    \caption{Percent difference in the turbulence with and without streaming at different redshifts. The percent difference generally declines as a function of redshift as the stream velocity decays. The turnover at $M \sim 10^6 \ \mathrm{M_\odot}$ also shifts slightly toward higher masses at earlier times and is marked by the gray shaded band.}
    \label{fig:diff}
\end{figure}

\begin{figure}[htbp]
    \centering
    \includegraphics[width=\columnwidth]{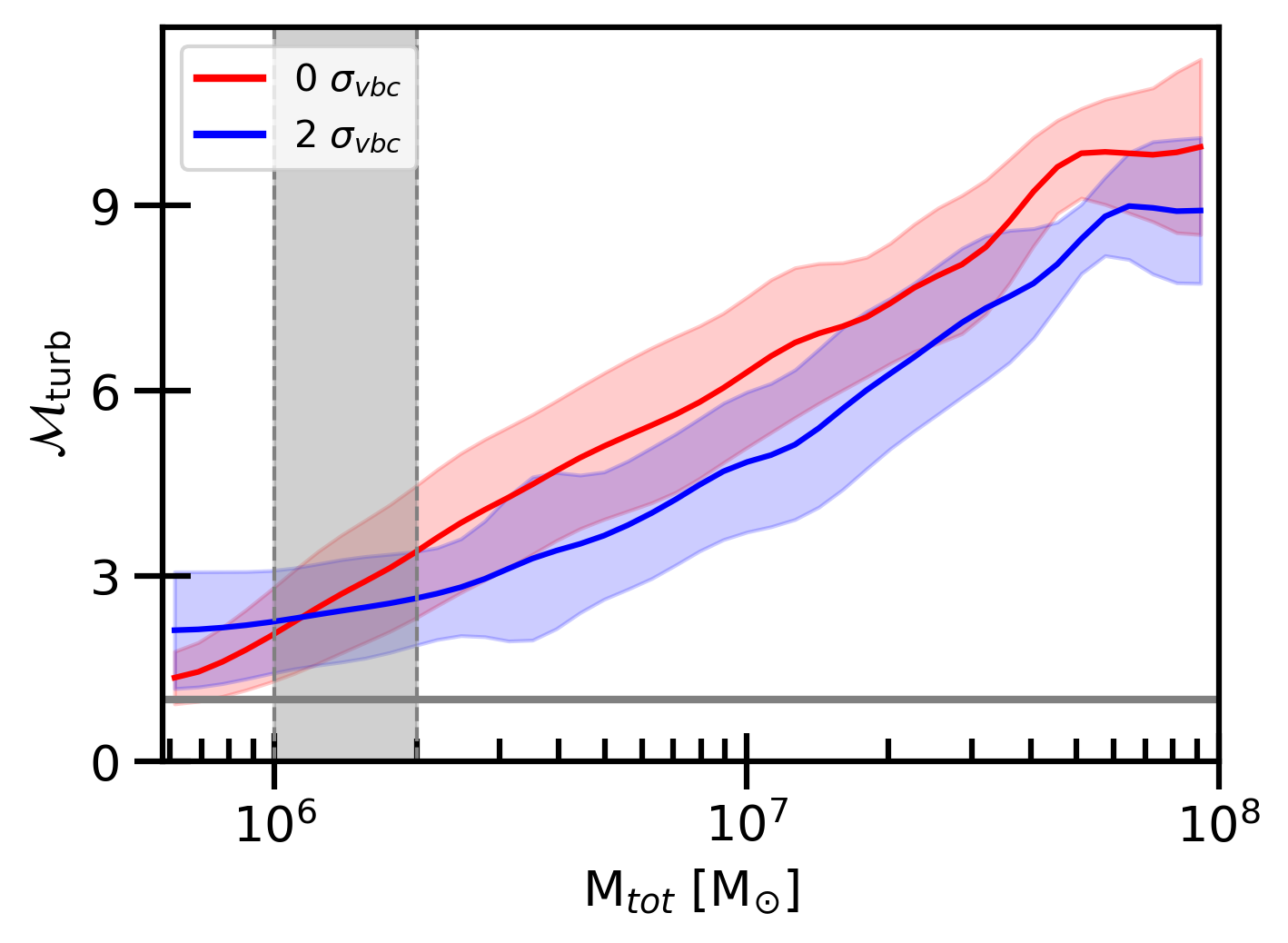}
    \caption{Turbulent Mach number as a function of total halo mass with and without streaming at $z = 20$. Shaded color bands indicate the $1\sigma$ scatter around the mean values. A clear turnover appears near $M \sim 10^6 \ \mathrm{M_\odot}$, marked by a gray band, mirroring the trend in the turbulent velocity (\autoref{fig:vel_comp}). Horizontal gray line marks a Mach number of 1.}
    \label{fig:turb}
\end{figure}

In \autoref{fig:vel_comp}, we show the mean turbulent, RMS, rotational, and radial velocity as a function of total halo mass for both the streaming ($v_{\rm bc} = 2\sigma_{\rm vbc}$, blue curve) and no-streaming ($v_{\rm bc} = 0\sigma_{\rm vbc}$, red curve) runs. In the top panel we see that streaming enhances turbulent velocity in low-mass halos ($M \lesssim 10^6 \ \mathrm{M_\odot}$) while suppressing it in high-mass halos ($M \gtrsim 10^6 \ \mathrm{M_\odot}$). To confirm the robustness of this turnover, we perform a higher-resolution simulation (see \autoref{app:resolution}) and find consistent results, including a clear turnover around $M \sim 10^6 \ \mathrm{M_\odot}$.

We also observe a strong correspondence between the RMS (second panel from top) and turbulent velocity curves across both runs: the difference between the streaming and the no-streaming run is similar in magnitude and exhibits a turnover at approximately the same mass scale. This consistency is expected, as the turbulent velocity is computed from the RMS velocity after subtracting coherent motions.

In the third panel of \autoref{fig:vel_comp}, we show the rotational velocity curves. We see that streaming enhances rotational velocities across the entire mass range, consistent with prior studies that showed streaming increases halo spin parameters \citep[e.g.,][]{Chiou+18, Druschke+18, Williams+23}. This effect likely arises from the coherent displacement of baryons relative to the halo center, which increases the net angular momentum of the system \citep{Druschke+18}.

To assess whether the turnover in the turbulent velocities reflects a statistically significant difference in the underlying distributions, we performed a two-sample Kolmogorov–Smirnov (KS) test on the turbulence values in the $v_{\rm bc} =0\sigma_{\rm vbc}$ and $v_{\rm bc} =2\sigma_{\rm vbc}$ runs, divided at the characteristic mass scale of $M =10^6 \ \mathrm{M_\odot}$. For halos with  $M < 10^6 \ \mathrm{M_\odot}$, we find a KS statistic of 0.35 with a p-value of $ \sim 10^{-67}$, indicating a highly significant difference between the distributions. For halos with $M \geq 10^6 \ \mathrm{M_\odot}$, the KS statistic is 0.10 with a p-value of $\sim 10^{-9}$, demonstrating that the difference remains statistically significant, though less pronounced. These results confirm that the turnover in turbulence is not only visually apparent but also statistically robust.

The difference in average turbulent velocities between the streaming and no-streaming runs is most pronounced at high redshift and in low-mass halos, and generally decreases with time. In \autoref{fig:diff}, we show the percent difference in turbulent velocity as a function of total halo mass and redshift, computed separately for each mass bin. The turnover remains near $M \sim 10^6 \ \mathrm{M_\odot}$ across all redshifts with a modest shift toward higher masses at earlier times. The percent difference is largest for low-mass halos, reaching factors of $\sim 1.5$, and declines steeply with increasing mass.  However, in higher-resolution runs, the percent difference appears to plateau at the lowest halo masses for the $v_{\rm bc} =2\sigma_{\rm vbc}$ run, indicating a potential saturation in the streaming-induced boost (see \autoref{app:resolution}). Future simulations with even higher resolution will be necessary to more precisely characterize the mass dependence of streaming-driven turbulence in halos with $M \lesssim 10^6~\mathrm{M_\odot}$. In \autoref{app:depen} we show the dependence on the turbulence when $v_{\rm bc} = 1\sigma_{\rm vbc}$ and  $v_{\rm bc} = 3\sigma_{\rm vbc}$.

The turbulent Mach number is a key diagnostic of the compressibility and energetics of turbulence in the halo. Following \cite{Greif_12}, we define it as:
\begin{equation}
\mathcal{M}_{\mathrm{turb}} = \frac{v_{\mathrm{turb}}}{\langle c_s \rangle} \ ,
\end{equation}
where \( \langle c_s \rangle \) is the mass-weighted sound speed and \( v_{\mathrm{turb}} \) is given by \autoref{eq:turbulent_velocity}. In \autoref{fig:turb}, we show the turbulent Mach number as a function of total halo mass. We see the clear bifurcation across $M \sim 10^6 \ \mathrm{M_\odot}$, analogous to the trend observed in the turbulent velocity. In low-mass halos ($M \lesssim 10^6 \ \mathrm{M_\odot}$), the turbulent Mach number is significantly enhanced in the streaming run, with boosts similar to that seen in the turbulent velocity. For halos of $M \sim 10^7 \, \mathrm{M_\odot}$ we find supersonic turbulence, with  $\mathcal{M}_{\mathrm{turb}} \sim 6$ in the no-streaming run consistent with the values recently reported by \cite{Chen_25} for similarly sized halos at comparable redshifts.

\subsection{Possible Sources of the Turbulence}
\label{subsec:exploring}

The enhancement of turbulence in low-mass halos (i.e., $M \lesssim 10^6 \ \mathrm{M_\odot}$) in the presence of a nonzero stream velocity is not surprising. At the time of recombination, the relative velocity between baryons and DM introduces coherent bulk motions into the gas. In our simulations, this is modeled as a uniform boost in the positive $\hat{x}$-direction. These bulk motions inject additional kinetic energy into the gas, which acts as an effective pressure that continuously suppresses density fluctuations on all scales. It also gets converted into radial and azimuthal motions along with turbulence inside low-mass DM halos through nonlinear interactions and the churning of the gas. These resulting velocities are dynamically important as their magnitudes are comparable to the sound speed in these halos.

\begin{figure}[htbp]
    \centering
    \includegraphics[width=\columnwidth]{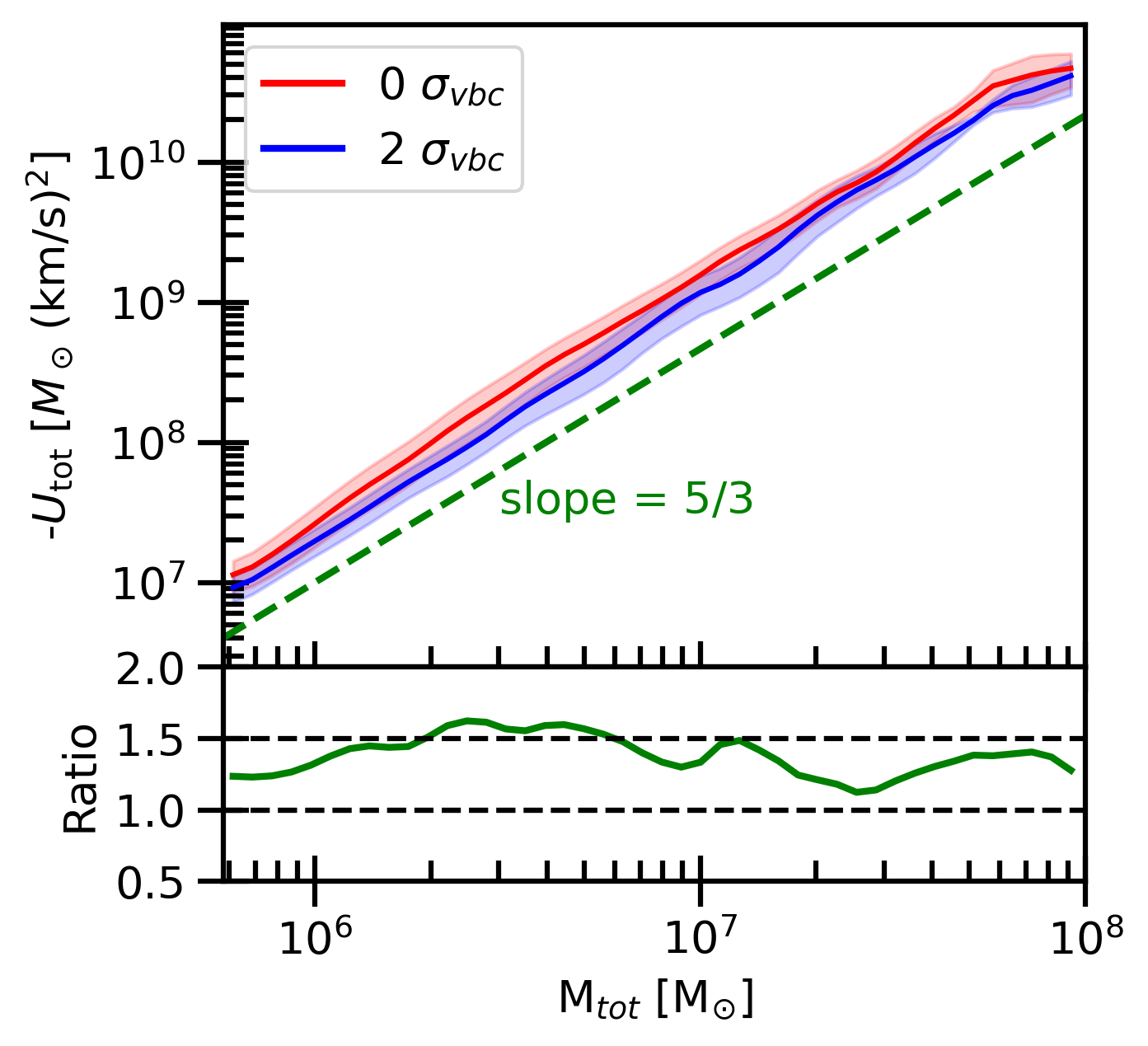}
    \caption{Gravitational potential energy of halos as a function of total mass (gas + DM) at $z = 20$ with and without streaming. The top panel shows the halo's gravitational potential energy, computed from gas--gas, DM--DM, and gas--DM interactions for all gas cells and dark matter particles within each halo. The green dashed line shows the expected scaling of slope \(5/3\) for virialized halos. Shaded color bands indicate the $1\sigma$ scatter around the mean values. The bottom panel shows the ratio of the two curves. No-streaming halos exhibit systematically higher potential energies across the full mass range.}
    \label{fig:potentials}
\end{figure}

The transition in turbulence near $M \sim 10^6 \ \mathrm{M_\odot}$ may result from a shift in the forces shaping the gas dynamics within these halos. In the absence of stellar feedback, the interplay between gravity and internal pressure, including thermal pressure and effective pressure from the stream velocity, determines the dynamics of the gas. As shown by \cite{Tes+10b}, the filtering mass, a cumulative analog to the Jeans mass that accounts for the time-dependent suppression of baryonic fluctuations \citep[e.g.,][]{Gnedin00,Naoz+07}, is approximately $1.07 \times 10^6 \ \mathrm{M_\odot}$ at $z = 20$ for $v_{\rm bc} =2\sigma_{v_{\mathrm{bc}}}$ \citep[see  Table~1 in][]{Tes+10b}. This value closely matches the mass scale at which turbulence transitions from being enhanced to suppressed. Below the filtering mass, pressure forces, including those introduced by streaming, dominate and inhibit gas collapse, while above it, gravitational potential wells are sufficiently deep that pressure support becomes subdominant. This scale-dependent behavior is consistent with previous studies that showed the stream velocity has its greatest dynamical impact on low-mass halos by reducing the gas fraction, delaying baryonic collapse, and suppressing the formation of small-scale structure. \citep[e.g.,][]{Greif10, Naoz+11a, Naoz+12,Popa+15,Chiou+18}

For halos with $M \gtrsim 10^6 \ \mathrm{M_\odot}$, the suppression of turbulence in the presence of streaming may at first seem counterintuitive. While the energy introduced by the stream velocity is expected to be less significant in high-mass halos, it is not immediately obvious why it would lead to a reduction in turbulence. A plausible explanation lies in an alternative mechanism of generating turbulence in primordial halos: accretion flows {\citep[e.g.,][]{Wise07, Greif08}}. As we will see, the stream velocity modifies the density field and accretion patterns, likely reducing the effectiveness of accretion-driven turbulence in high-mass halos.

To explore the possibility that the stream velocity suppresses accretion-driven turbulence in high-mass halos, we begin by analyzing the radial speed of incoming gas. In the bottom panel of \autoref{fig:vel_comp}, we show the mass-weighted RMS radial velocity magnitude of gas within halos as a function of total mass. For halos $M \gtrsim 10^6 \ \mathrm{M_\odot}$ we see a slight increase in the kinetic energy associated with radial motions without streaming. The gas inside the halos is also denser without streaming {\citep[e.g.,][]{OLMc12, Richardson+13}}. Taken together, this suggests that the energy of accretion flows will be higher in high-mass halos without streaming. This inflowing accreting gas can generate turbulence  through shear flows, instabilities, or the thermalization of kinetic energy via shocks. We explore the latter phenomenon in \autoref{subsec:thermalization}.

\begin{figure*}
    \centering
    \begin{minipage}{\textwidth}
        \includegraphics[width=\textwidth]{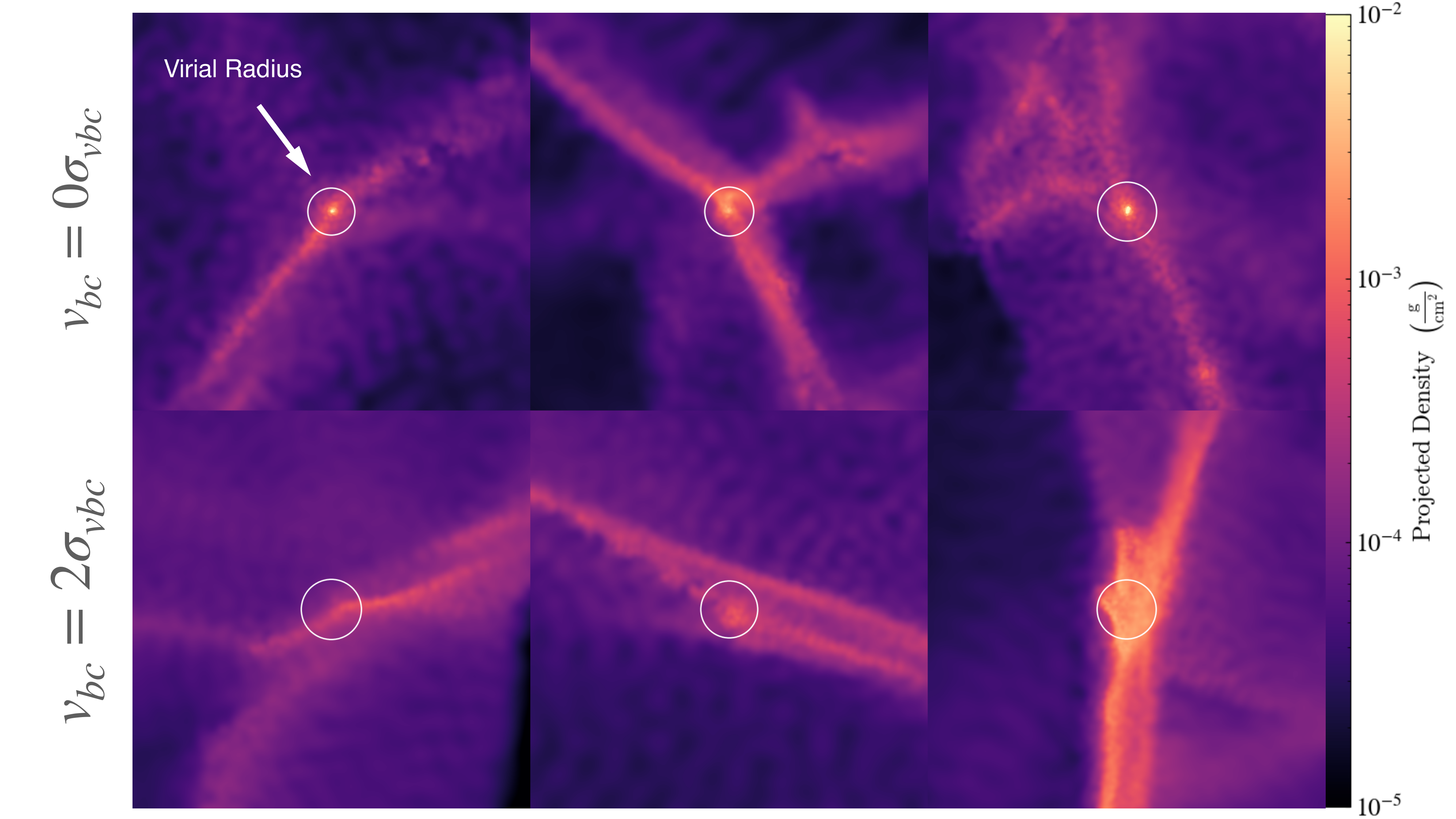}
    \end{minipage}
    \begin{minipage}{\textwidth}
        \includegraphics[width=\textwidth]{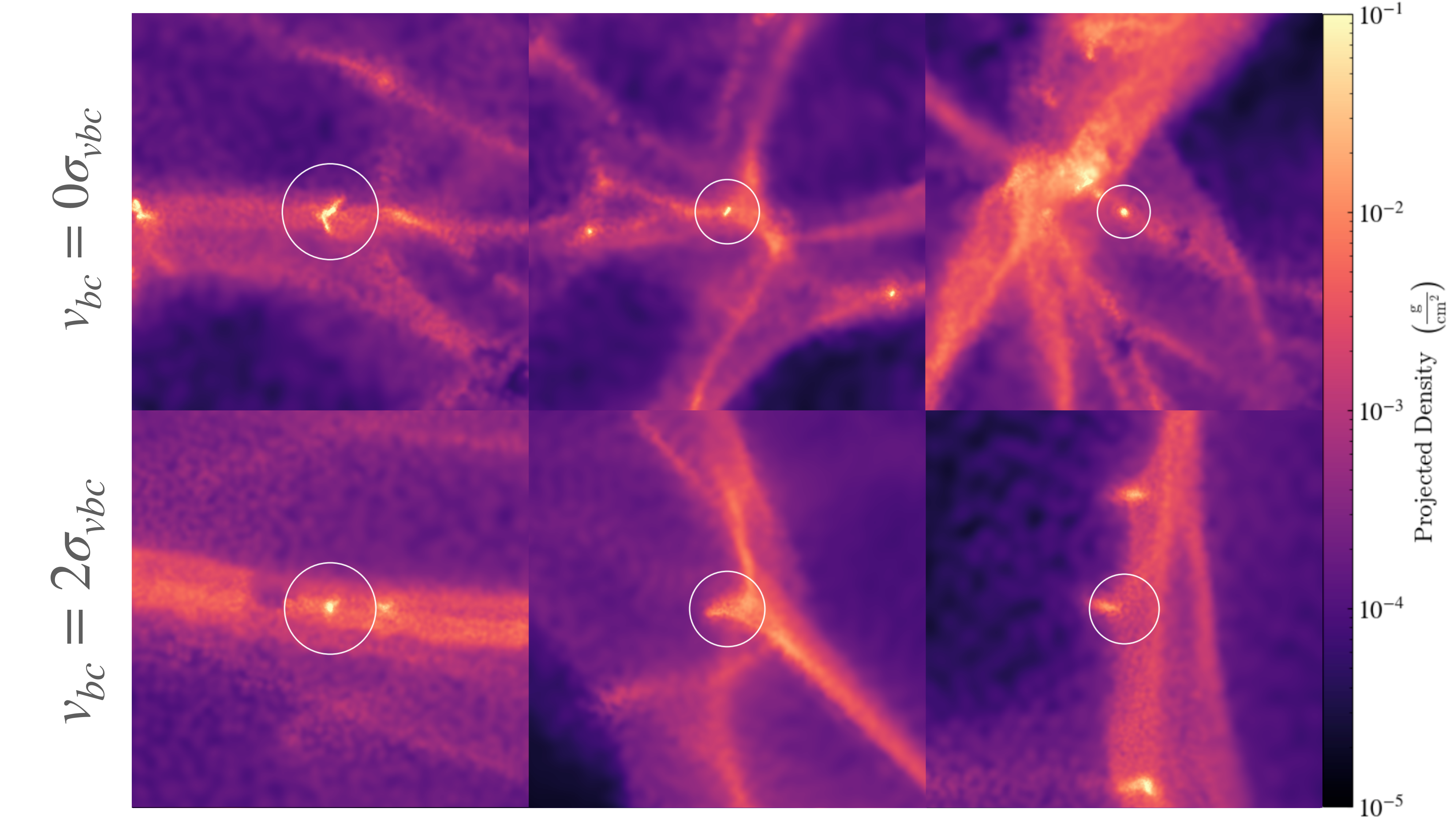}
    \end{minipage}
    \caption{Projected gas density surrounding representative halos from two mass bins in simulations with streaming (bottom rows in each pair) and without streaming (top rows in each pair) at $z = 20$. Each panel shows a 1 kpc (physical) cutout centered on the halo's center of mass. The top pair depict halos with $ M < 10^6\ \mathrm{M_\odot}$, while the bottom pair depict halos in the range $M =10^6 - 10^7 \ \mathrm{M_\odot}$. White circles denote the virial radius of each halo. Low-mass halos appear similarly isolated in both streaming and no-streaming runs and are embedded along thin filaments. In contrast, higher-mass halos without streaming reside in denser environments with more frequent filamentary intersections and nearby structures, conditions favorable for mergers. This visual difference suggests that streaming may suppress merger-driven turbulence in high-mass halos by reducing the abundance of dense nearby structures.}
    \label{fig:halos}
\end{figure*}

One possible explanation for the elevated radial velocities observed in more massive no-streaming halos is their deeper gravitational potential wells \citep[e.g.,][]{Williams+25}. We can compute the total gravitational potential energy of each halo by summing three contributions: gas–gas self-interaction, DM–DM self-interaction, and gas–DM cross-interaction. These quantities are calculated using all gas cells and DM particles associated with each halo. The total potential energy is given by:
\begin{equation}
    U_{\text{total}} = U_{\text{gas-gas}} + U_{\text{dm-dm}} + U_{\text{gas-dm}} \,.
\end{equation}
 Each term is calculated via pairwise summation over cell and particle pairs:
\begin{align}
    U_{\text{gas-gas}} &= -\frac{1}{2} G \sum_{i \neq j}^{N_{\text{gas}}} \frac{m_i m_j}{|\mathbf{r}_i - \mathbf{r}_j|} \,, \\
    U_{\text{dm-dm}}   &= -\frac{1}{2} G \sum_{i \neq j}^{N_{\text{dm}}} \frac{m_i m_j}{|\mathbf{r}_i - \mathbf{r}_j|} \,, \\
    U_{\text{gas-dm}}  &= - G \sum_{i=1}^{N_{\text{gas}}} \sum_{j=1}^{N_{\text{dm}}} \frac{m_i m_j}{|\mathbf{r}_i - \mathbf{r}_j|} \,.
\end{align}
 where \( m_i \) and \( \mathbf{r}_i \) denote the mass and position of particle \( i \), and \( G \) is the gravitational constant in units of \(\mathrm{kpc} \, (\mathrm{km/s})^2 \, M_\odot^{-1}\). In \autoref{fig:potentials}, we plot the total gravitational potential energy as a function of halo mass. We find that halos in no-streaming exhibit deeper potential wells compared to their streaming counterparts. This is expected as the stream velocity makes it harder for potential wells to form by adding an effective anisotropic pressure to the gas. This is the same underlying physics that causes a suppression of the gas density (in halos of all sizes) and the decrease in the number of DM halos. For a fixed halo mass, a deeper gravitational potential would lead to greater radial energies in accreting flows. These larger radial velocities would amplify turbulence and could help explain the trends seen in \autoref{fig:vel_comp} and \autoref{fig:diff}.

\begin{figure*}[htbp]
    \centering
    \includegraphics[width=0.49\textwidth]{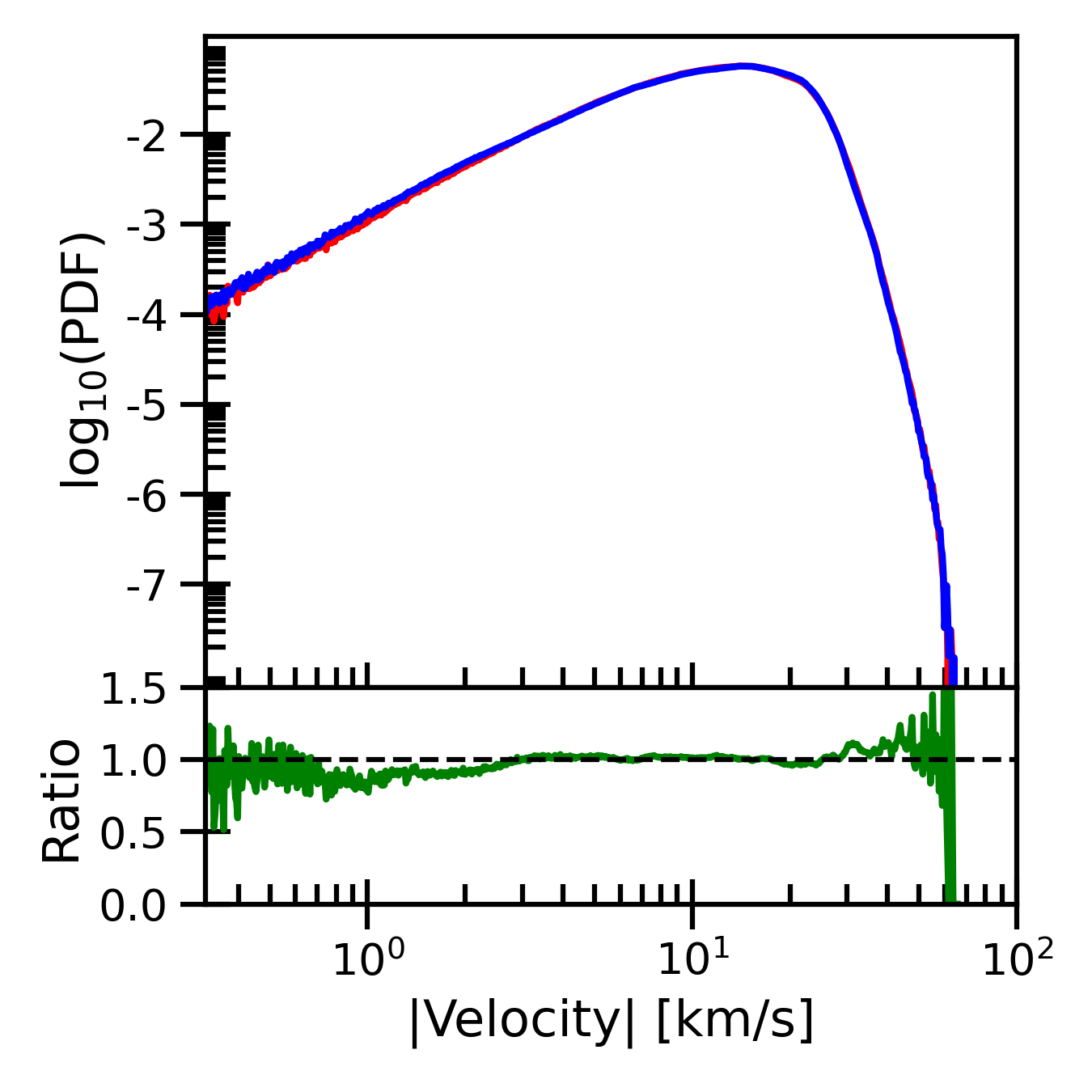}
    \includegraphics[width=0.49\textwidth]{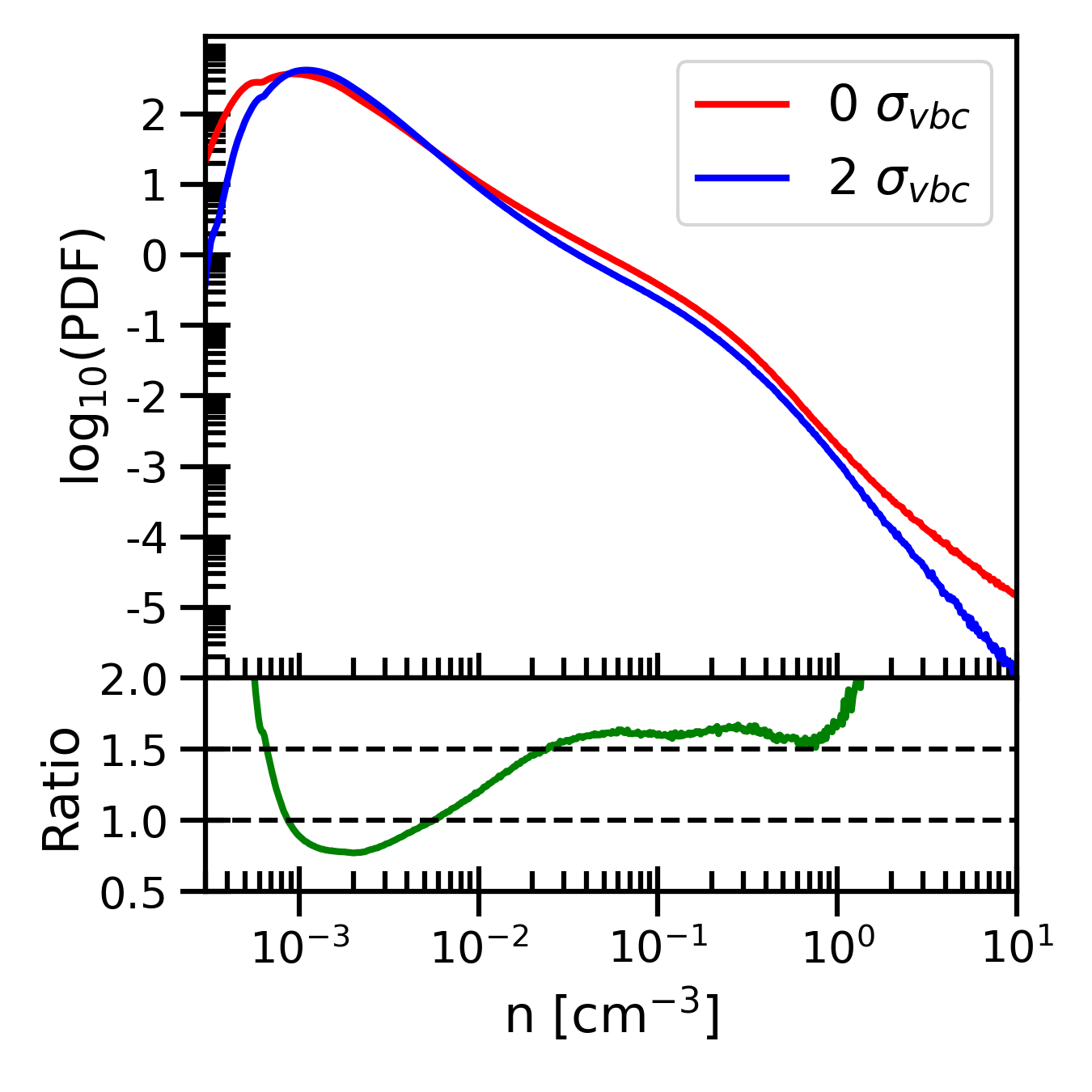}
    \caption{Probability density functions (PDFs) of gas velocity (left) and number density (right) in the intergalactic medium (IGM) at $z = 20$ with and without streaming. The IGM is defined as all gas not inside halos as identified by the FoF. Bottom subpanels show the ratio of the PDFs in each run. While the velocity distributions are nearly identical, indicating that the kinematic imprint of streaming has largely dissipated by this epoch, the number density distribution reveals a suppression of dense filamentary gas ($n \gtrsim 10^{-1} \ \mathrm{cm^{-3}}$) in the streaming run. This difference, integrated over cosmic time, contributes to reduced kinetic energy and turbulence in gas accreting onto halos in the presence of streaming.}
    \label{fig:pdf_igm}
\end{figure*}

Mergers are also known to generate turbulence inside primordial halos \citep[e.g.,][]{Wise07, Prieto_12, Bovino_14}. They also raise the temperatures of the gas \citep[e.g.,][]{Yoshida2003}, which would increase the accretion rate of the halo \citep[e.g.,][]{Hirano+15}. Although the impact of the stream velocity on mergers has not yet been studied, it is reasonable to expect a suppression, given that streaming suppresses structure formation and reduces the overall abundance of DM halos \citep[e.g.,][]{Tes+10a, Naoz+11a}. This reduction in mergers may contribute to the observed suppression of turbulence in high-mass halos when streaming is present.

To qualitatively explore this possibility, we look at the environments of typical halos above and below the turnover of $M \sim 10^6 \ \mathrm{M_\odot}$. In the top panel of \autoref{fig:halos} we show three typical halos with $M <10^6 \ \mathrm{M_\odot}$ for streaming and no-streaming in a 1 kpc cutout centered on the center of mass of the selected halo, and in the bottom panel we show three typical halos with $M = 10^6 - 10^7 \ \mathrm{M_\odot}$ at $z = 20$. For both streaming and no-streaming we see that halos with $M <10^6 \ \mathrm{M_\odot}$ generally exist in isolation, residing along wisps of cosmic filaments. On the other hand for halos with $M = 10^6 - 10^7 \ \mathrm{M_\odot}$ we see that the no-streaming halos now exist in a more complex environment, with more nearby structures and multiple knots of cosmic filaments. The streaming halos, on the other hand, still form in less crowded environments, with fewer structures and fewer intersected cosmic filaments. Thus, it is more likely that mergers, the likelihood of which increases with halo mass, is preferentially increasing the turbulence for high-mass halos in the no-streaming run. 

\subsection{The Intergalactic Medium}
\label{subsec:IGM}

We next look at the properties of the intergalactic medium (IGM) at $z = 20$ to further explore whether the enhancement in turbulence in the no-streaming high-mass halos is related to the turbulence generated during accretion. In \autoref{fig:pdf_igm}, we show the probability density functions (PDFs) of gas velocity (left panel) and number density (right panel) for all gas not inside DM halos, as defined in \autoref{subsection:cosmological}. The bottom panels in each figure show the corresponding ratios between the curves.

The velocity PDFs indicate that by at least this epoch, the stream velocity's kinematic imprint is no longer prominent in the IGM. The bulk of the gas exhibits velocities well in excess of the decayed streaming value, and both simulations yield nearly identical velocity distributions for gas not inside halos. This suggests that the residual effect of the stream velocity on the velocities in the IGM is negligible, consistent with what was shown earlier that the stream velocity's influence persists primarily in halos with $M \lesssim 10^6 \ \mathrm{M_\odot}$.

The number density PDF reveals a more pronounced difference. For the most diffuse gas ($n \lesssim 10^{-4} \ \mathrm{cm}^{-3}$), which constitutes the majority of the IGM by volume, the density distribution is similar in both runs. In the cosmic filaments, however, which have typical densities \( n \gtrsim 10^{-1} \, \mathrm{cm}^{-3} \), streaming suppresses the gas densities. This suppression is a cumulative effect over cosmic time and is consistent with the diffuse gas observed in halos. When this denser gas accretes into the halos, it will have higher kinetic energy and will therefore generate more turbulence during accretion. It will also thermalize more of its energy, something we explore in the next section. Despite the suppression of gas densities with streaming, however, mass conservation is maintained, as the displaced gas is redistributed into lower-density regions of the IGM. 

\subsection{Thermalization of Accretion Shocks}
\label{subsec:thermalization}

As discussed in the previous section, the enhanced turbulent velocities in high-mass halos without streaming may be linked to accretion-driven turbulence. To further explore this possibility, we analyze the thermalization of accretion-driven shocks using the shock finder developed by \citet{Schaal2015, Schaal_2016}. The shock finder loops over all gas cells, identifies those that have been shocked\footnote{Here we summarize the shock finder principles:  it first identifies a \textit{shock zone} comprising cells that satisfy three criteria: (i) the flow is compressive, $\nabla \cdot \vec{v} < 0$; (ii) the temperature and density gradients are aligned, $\nabla T \cdot \nabla \rho > 0$; and (iii) the logarithmic jumps in temperature and pressure each exceed a threshold corresponding to a minimum Mach number of 1.3, to avoid spurious shocks. From each shock zone cell, rays are cast along the direction of increasing temperature (i.e., toward the post-shock region) and then reversed toward the pre-shock region. The cell with the most negative velocity divergence (i.e., maximum compression) along this path is tagged as part of the \textit{shock surface}. These \textit{shock surface} cells are the ones considered shocked; see Figure~1 of \citet{Schaal2015} for an illustration.} and computes the energy dissipation rate along with the Mach number for each shocked cell. The energy dissipation rate of the shock, the quantity of interest, is computed by dividing the thermal energy flux by the surface of the shocked Voronoi cell. The thermal energy flux is given by:
\begin{equation}
f_\text{th}=\delta(\Machnum)f_\Phi ,
\label{eq:f_thermal}
\end{equation}
 where $f_{\Phi}$ is the incoming kinetic energy and $\delta(\Machnum)$ is the thermalization efficiency. The incoming kinetic energy is given by:
 \begin{equation}
 f_\Phi = \frac{1}{2}\rho_1(c_1\Machnum)^3 ,
 \end{equation}
 and the thermalization efficiency is given by \citep{Kang_07}:
\begin{equation}
\delta(\Machnum)=\frac{2}{\gamma(\gamma-1)\Machnum^2R}\left[\frac{2\gamma\Machnum^2-(\gamma-1)}{(\gamma+1)}-R^\gamma\right] ,
\label{eq:delta_M}
\end{equation}
where R is the density jump given by:
\begin{equation}
R\equiv\frac{\rho_2}{\rho_1}=\frac{(\gamma+1)\Machnum^2}{(\gamma-1)\Machnum^2+2} .
\label{eq:R}
\end{equation}

\begin{figure*}[ht!]
    \centering
    \includegraphics[width=\textwidth]{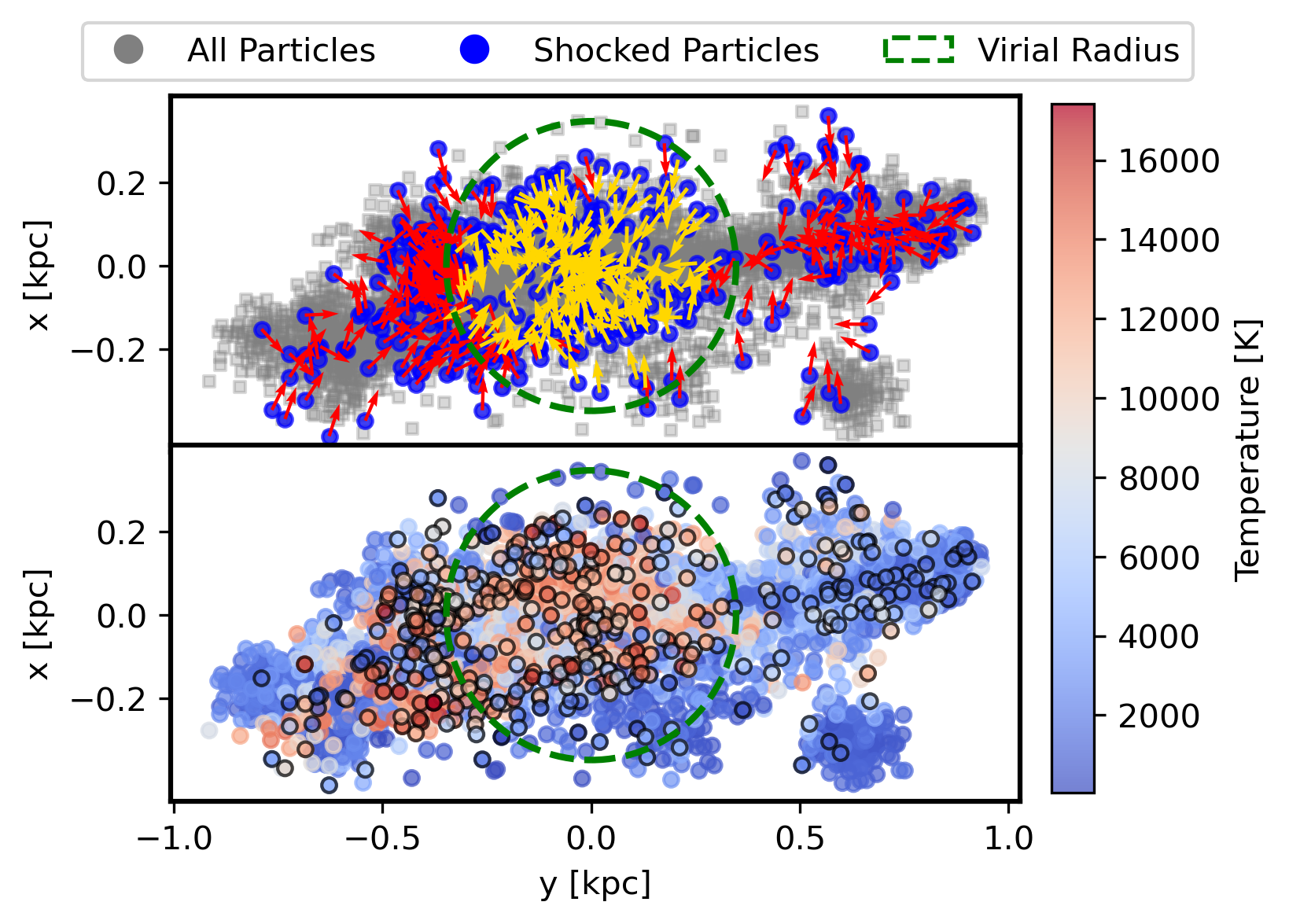}
    \caption{Shocked gas in an illustrative halo of mass $M \simeq 10^7 \ \mathrm{M_\odot}$ at $z = 20$. \textbf{Top panel:} All gas within the halo are shown in gray, while shocked gas are overplotted in blue. Red arrows indicate normalized velocity vectors of shocked gas outside the virial radius of the halo, and yellow arrows indicate those inside. The dashed green circle marks the virial radius. We see that the shocked gas predominantly traces radially oriented accretion shocks. \textbf{Bottom panel:} Gas temperature distribution, where shocked gas are outlined in black. Shocked gas reaches temperatures near the virial temperature, consistent with strong thermalization during accretion.}
    \label{fig:shock}
\end{figure*}

In \autoref{fig:shock} we show the distribution of shocks for a no-streaming halo of mass $M \simeq 10^7 \ \mathrm{M_\odot}$. In the top panel, we show the shocked gas cells in blue along with the virial radius of the halo represented by green circles overlaid with normalized velocity vectors, colored red for those outside the virial radius and yellow for those inside. In this particular halo, shocked gas is found both inside and outside the virial radius. The bottom panel shows the corresponding temperature distribution of the gas, with shocked cells marked by black circles. We find that some of the shocked gas reaches temperatures comparable to the virial temperature of the halo, which we compute as:
\begin{equation}
T_{\text{vir}} = \frac{\mu m_p}{2 k_B} \frac{G M_{\text{vir}}}{R_{\text{vir}}} ,
\end{equation}
 where $\mu = 1.22$ is the mean molecular weight for neutral primordial gas, $m_p$ is the proton mass, $k_B$ is the Boltzmann constant, and $G$ is the gravitational constant. For this halo, we find $T_{\text{vir}} = 1.12 \times 10^4\ \mathrm{K}$.

\begin{figure}[htbp]
    \centering
    \includegraphics[width=\columnwidth]{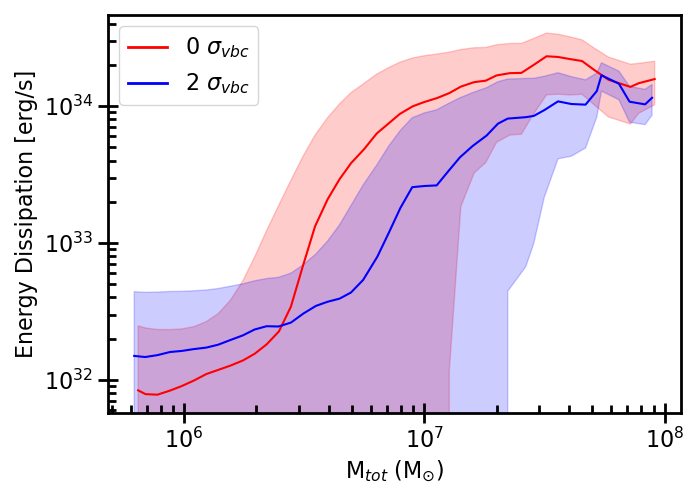}
    \caption{Average energy dissipation rate from shocked gas as a function of total halo mass at $z = 20$. Shaded color bands indicate the $1\sigma$ scatter around the mean values. At high-mass halos ($M \gtrsim 10^6 \ \mathrm{M_\odot}$), no-streaming halos exhibit higher dissipation rates, consistent with enhanced energy in accretion flows.}
    \label{fig:edis}
\end{figure}

The measured energy dissipation from shocked gas provides further support that streaming may suppress accretion-driven turbulence in high-mass halos. In \autoref{fig:edis} we show the average energy dissipation rate of shocked gas inside halos as a function of halo mass. We see that for halos with $M \gtrsim 2 \times 10^6 \ \mathrm{M_\odot}$, no-streaming halos exhibit systematically higher energy dissipation. When this energy is thermalized in shocks, some of it can transform into non-thermal motions through post-shock Richtmyer–Meshkov and Kelvin–Helmholtz instabilities, which will stir the gas and generate turbulence. Thus, no-streaming halos, which have a higher energy dissipation rate, will also have higher turbulence associated with these shocks. In \autoref{fig:edis} we also see that the energy dissipation rate turnovers at around the same mass scale as the turbulence (i.e $M \sim 10^6 \ \mathrm{M_\odot}$), further suggesting a causal link.  

The enhanced energy dissipation observed in high-mass halos without streaming can be partly attributed to the denser infalling intergalactic gas. As discussed in \autoref{subsec:IGM} and shown in \autoref{fig:pdf_igm}, filamentary structures in the no-streaming case exhibit systematically higher densities. When this denser material accretes onto halos, it carries greater kinetic energy, leading to an elevated thermal energy flux upon shocking as described by \autoref{eq:f_thermal}. The thermalization efficiency, however, also depends on the Mach number, which, though saturating to a max value of 0.55 for $\gamma = 5/3$ when $\mathcal{M} \approx 10$, will also play a role in determining the energy dissipation rate. Combined with the higher radial velocities shown in \autoref{fig:vel_comp}, these denser inflows likely contribute to stronger shock heating and enhanced accretion-driven turbulence in high-mass halos in the no-streaming run.

\section{Discussion}
\label{sec:discussion}

The enhancement of turbulence and rotational velocities in low-mass halos in the presence of streaming is consistent with prior works. \citet{Greif+11} found that for a $v_{\rm bc} = 1\sigma_{\rm vbc}$ turbulence is increased in several minihalos ($M \lesssim 10^6 \ \mathrm{M_\odot}$) during runaway collapse, defined by central densities reaching $n_\mathrm{H} = 10^9 \ \mathrm{cm^{-3}}$. Although halos in our cosmological simulation do not reach these number densities, we still observe enhanced turbulence for halos in a similar mass range. This suggests that turbulent kinetic energy present at the halo scale cascades to smaller scales, a process also known to occur in local molecular clouds \citep[e.g.,][]{Larson_1981}. The higher rotational velocities we find in the streaming runs is likewise consistent with the elevated spin parameters generated by streaming \citep[e.g.,][]{Chiou+18, Druschke+18, Williams+23}.

The stream velocity may also influence the Pop III initial mass function (IMF) in a halo mass-dependent manner. If star formation first occurs in halos with $M \lesssim 10^6 \ \mathrm{M_\odot}$, then our results and those of \cite{Greif+11} suggest that streaming-induced turbulence is enhanced on both large and small scales. This enhanced turbulence in low-mass halos, which may be amplified during gravitational contraction \citep[e.g.,][]{Higashi2021}, may increase fragmentation, and may be contributing to the increased star formation rates found in low-mass objects in \cite{Lake+24a}. 

Streaming is also known to delay the onset of Pop III star formation and to raise the halo mass threshold at which collapse occurs \citep[e.g.,][]{Stacy+10, Maio+11, Hirano+18, Kulkarni_21...917...40K}. These more massive halos typically collapse more rapidly and tend to form a greater number and more massive stars \citep[e.g.,][]{Li_24, hirano_25}. If star formation primarily occurs in halos with $M \gtrsim 10^6 \ \mathrm{M_\odot}$, as suggested by the critical halo mass threshold, $M_\mathrm{crit}$ \citep[e.g.,][]{Kulkarni_21...917...40K, Schauer_21, Hegde_23, Nebrin_23}, then our results indicate that streaming suppresses turbulence relative to no-streaming halos of similar halo mass at the same redshift (\autoref{fig:vel_comp}). At fixed halo mass, the differing levels of turbulence may lead to variations in the IMF, as turbulence is known to promote fragmentation both at the cloud scale \citep{Sugimura_2023} and within protostellar disks \citep[e.g.,][]{Clark_08A, Wollenberg2020, Riaz23}. This, of course, does not account for feedback from the differing star formation histories between streaming and no streaming. This critical mass threshold can also be modulated by additional backgrounds, increasing in the presence of a uniform Lyman-Werner field \citep[e.g.,][]{Kulkarni_21...917...40K, Schauer_21}, or decreasing with a uniform X-ray background \citep[e.g.,][]{Hummel_15, Park_21} or under the influence of cosmic rays \citep{Hummel_16}. 

Streaming enhancement of rotational motions and halo spin could also impact the number and masses of protostars. Enhanced rotational motion in collapsing clouds generally inhibits fragmentation of the protostellar disk and the mass accretion onto individual protostars \citep[e.g.,][]{Clark11b, Hirano+14, Dutta_16, Riaz18, Jaura_22, Wollenberg2020}. Although the scales we study are larger than those examined in core-scale simulations, conservation of angular momentum suggests that enhanced large-scale rotation may cascade inward. The elevated spin parameters seen in streaming runs may even impact the orbital structure of protostars \citep{Sugimura_2023}. How this enhancement in rotational motions couples with the turbulent velocities to impact the IMF will require further high-resolution simulations. 

The accretion shocks identified in our simulations are similar to those reported in recent high-resolution studies of primordial halos. \cite{Kiyuna_23} showed that accreting gas onto halos of mass $5.0 \times 10^6\ \mathrm{M_\odot}$ experience shocks at around the virial radius at  $z = 20$. Similar results were also found in \cite{Fernandez_2014} \& \cite{Wise07}. Halos in our simulation also exhibit shock-heated gas near the virial radius, with some halos showing shocked gas extending into the halo interior as depicted in \autoref{fig:shock}. 

The classical framework for describing gas accretion distinguishes between hot-mode and cold-mode accretion. In this paradigm, low-mass halos accrete gas via cold streams along filaments that penetrate the halo without shock heating to the virial temperature, while in more massive halos, gas undergoes spherical accretion shocks and is heated to the virial temperature \citep{Keres_2005, Birnboim_2003}. The recent works of \cite{Kiyuna_23} and \cite{Fernandez_2014} have added subtleties to this paradigm in the high-redshift universe. The former by showing that the gas temperature along accretion columns is $ T\sim 10^4$ K, and the latter by not observing cold accretion in their halos even though they nominally satisfy the criterion of \cite{Birnboim_2003}. While in this study we do not closely follow the thermal history of individual gas streams to determine when and how cold-mode accretion occurs, we do find that at least some of the accreting gas is shock-heated near and inside the virial radius in low-mass halos. 

The differences in energy dissipation rates of shocks induced by the stream velocity may lead to observable consequences. \citet{OLMc12} found bow shocks surrounding DM halos in the presence of streaming, and concluded that these shocks, as well as structure formation shocks, with or without streaming, would not erase a strong 21 cm absorption signal. However, they noted the caveat that the fraction of kinetic energy thermalized in these shocks remained uncertain \citep{McOL12}. In our simulations, we find shocked gas reaching temperatures of 10$^4$ K, consistent with those reported by \cite{Fernandez_2014} and \cite{Kiyuna_23}. These shocks may accelerate electrons, leading to inverse Compton (IC) emission spanning radio to soft X-ray frequencies, potentially detectable with upcoming instruments \citep{LoebFurlanetto+13}. A study into whether the differences in shock energy dissipation observed in our simulations could produce an observable signature of the stream velocity is left for future work.

\section{Conclusion}
\label{sec:conclusion}

In this paper, we sought to explain, for the first time, how the stream velocity impacts turbulence across a range of halo masses in the high-redshift universe. In addition to previously recognized drivers of turbulence in primordial halos, such as accretion flows and halo mergers, we find that the stream velocity itself can be a significant source of turbulence in low-mass halos. Our primary results can be summarized as follows:
\begin{enumerate}

\item We identify a mass threshold of $M \sim 10^6 \ \mathrm{M_\odot}$ that separates two distinct regimes. In halos below this threshold, a stream velocity of v$_{\rm bc}$ =  2$\sigma_{\rm vbc}$ enhances turbulence, whereas in halos above the threshold, it suppresses it (\S \ref{subsec:turb}). The enhancement for low-mass halos is likely sourced by the increased kinetic energy introduced by the stream velocity, while the suppression in high-mass halos likely results from weaker accretion-driven turbulence in the presence of the stream velocity. 
 
\item At $z = 20$, the velocity distribution of intergalactic gas is nearly identical in simulations with and without streaming, indicating that the kinematic imprint of the stream velocity has largely dissipated from the IGM. However, the density PDF shows that filamentary structures feeding DM halos remain systematically denser in the no-streaming run (\S\ref{subsec:IGM}). Additionally, halos without streaming exhibit deeper gravitational potential wells across all masses and higher radial velocities in high-mass halos. Together, these trends imply that accretion flows onto high-mass halos are more energetic in the absence of streaming.

\item The average energy dissipation rate from shocks exhibits a similar mass-dependent trend as the turbulence, with a turnover near $M \sim 2 \times 10^6 \ \mathrm{M_\odot}$. At higher masses, halos without streaming show greater shock dissipation rates than their streaming counterparts (\S\ref{subsec:thermalization}), consistent with the enhanced turbulence seen in this regime. This correlation supports the argument that more energetic accretion flows, facilitated by deeper potential wells and denser filaments, contribute to stronger shock heating and potentially to greater turbulence in high-mass halos without streaming.

\end{enumerate}

The stream velocity thus influences gas dynamics in two opposing ways: it enhances turbulence in low-mass halos by introducing additional kinetic energy into the gas, and suppresses turbulence in high-mass halos by lowering their potentials and reducing gas density both inside and outside halos.  Which of these effects dominates depends on halo mass, with a transition occurring near $M \sim 10^6\ \mathrm{M_\odot}$. This mass-dependent behavior suggests that the star formation at a fixed DM halo mass may differ in the presence of the stream velocity and could be relevant for future  high-resolution, full-physics simulations seeking to model the first stars with the stream velocity.

\vspace{1em}
{\bf Acknowledgments.} A.C. would like to thank Ka Ho Yuen for inspiring discussions. A.C., C.E.W., W.L., B.B., S.N., F.M., and M.V. thank the support of NASA grant Nos. 80NSSC20K0500 (9- ATP19-0020) and 80NSSC24K0773 (ATP-23- ATP23-0149). B.B. acknowledges support from NSF grant AST-2009679. This research was also supported in part by the National Science Foundation under Grant No. NSF PHY-1748958. B.B. is grateful for generous support from the David and Lucile Packard Foundation, the Alfred P. Sloan Foundation, and the Flatiron Institute, which is funded by the Simons Foundation. C.E.W.  acknowledges the support of the National Science Foundation Graduate Research Fellowship, the University of California, Los Angeles (UCLA), the UCLA Center for Diverse Leadership in Science Fellowship, and the UCLA Mani L. Bhaumik Institute for Theoretical Physics Fellowship. This material is based upon work supported by the National Science Foundation Graduate Research Fellowship Program under Grant No. DGE-2034835. Any opinions, findings, conclusions, or recommendations expressed in this material are those of the author(s) and do not necessarily reflect the views of the National Science Foundation. S.N. thanks Howard and
Astrid Preston for their generous support. The simulations and computations in this paper were carried on the Rusty superecomputer at the Flatiron Institute.

\software{ {\tt AREPO} \citep{2020ApJS..248...32W}, matplotlib \citep{Matplotlib}, numpy \citep{numpy}, scipy \citep{SciPy}, and {\tt yt} \citep{Turk+11}.}

\bibliography{cosmo}

\appendix
\section{Resolution Study}
\label{app:resolution}

\begin{figure}[htbp]
    \centering
    \includegraphics[width=\columnwidth]{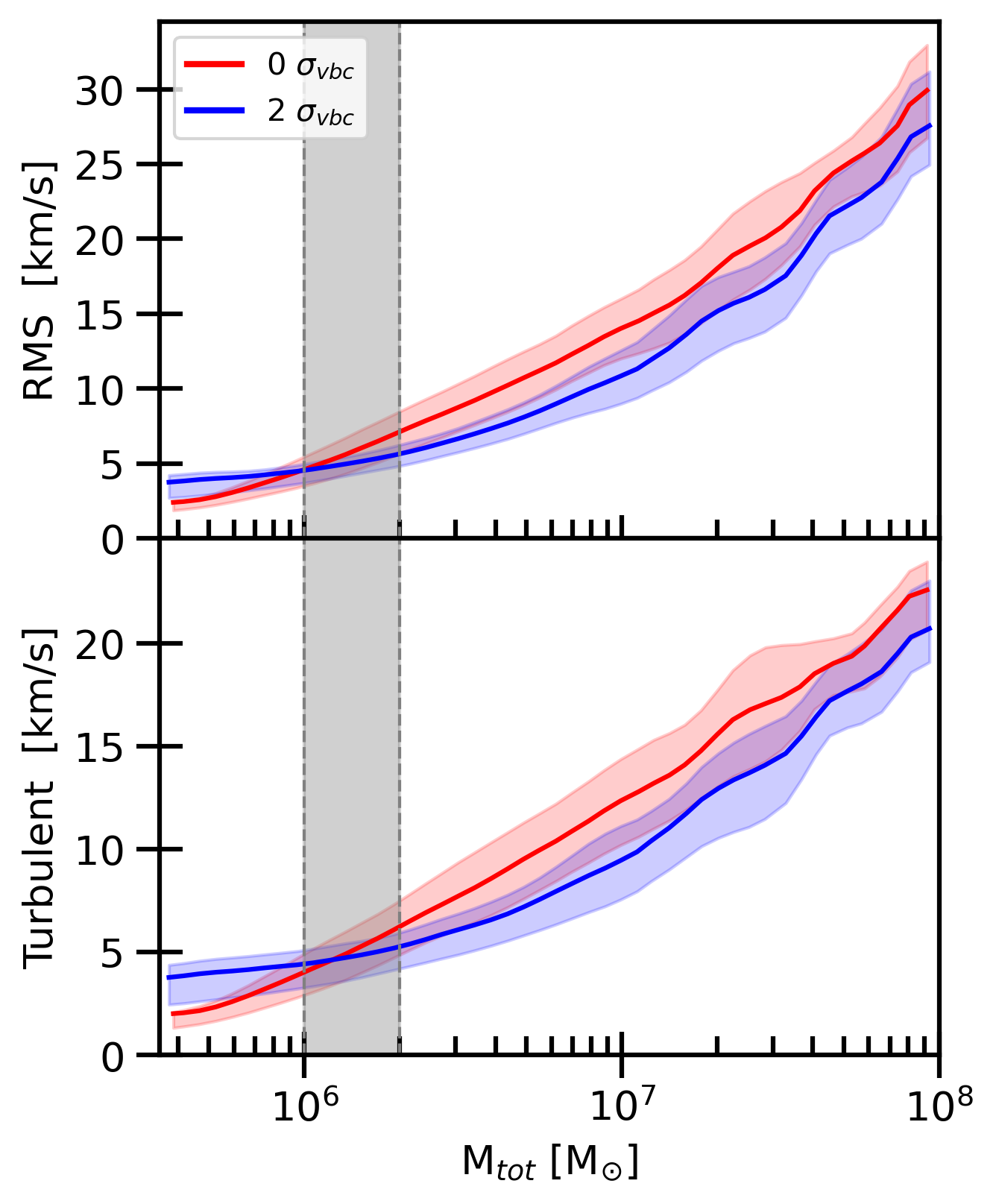}
    \caption{RMS and turbulent velocities as a function of halo mass in the higher-resolution run. Shaded color bands indicate the $1\sigma$ scatter around the mean values. The turnover in turbulence remains at $M \sim 10^6\ \mathrm{M_\odot}$, marked by the gray shaded band, while the difference between streaming and no-streaming begins to saturate for the lowest-mass halos.}
    \label{fig:768_velocities.png}
\end{figure}

To assess the impact of resolution on our velocity curves, we repeat the analysis using a higher-resolution setup: a 2.5 Mpc$^3$ box containing $768^3$ Voronoi gas cells (with initial gas mass $m_{\rm gas} = 200\ \mathrm{M_\odot}$) and $768^3$ dark matter particles (with initial mass $m_{\rm DM} = 1.1 \times 10^3\ \mathrm{M_\odot}$). In \autoref{fig:768_velocities.png}, we show the RMS and turbulent velocities from this simulation. The characteristic turnover in turbulent velocity between the streaming and no-streaming runs persists at $M \sim 10^6\ \mathrm{M_\odot}$. Additionally, the difference in turbulence appears to begin saturating at the lowest resolved halo masses. Even higher-resolution simulations capable of probing lower-mass halos will be necessary to determine whether this trend continues.

\section{Turbulence With Varying Stream Velocities}
\label{app:depen}

To see how the strength of the stream velocity influences turbulent velocities, we analyze simulations with $v_{\rm bc} = 1\sigma_{\rm vbc}$ and $v_{\rm bc} = 3\sigma_{\rm vbc}$, in addition to the $v_{\rm bc} = 2\sigma_{\rm vbc}$ run. For each case, we compute the percent difference in turbulent velocity relative to the corresponding no-streaming ($v_{\rm bc} = 0\sigma_{\rm vbc}$) simulation, separately in each mass bin.

\begin{figure}[htbp]
    \centering
    \includegraphics[width=\columnwidth]{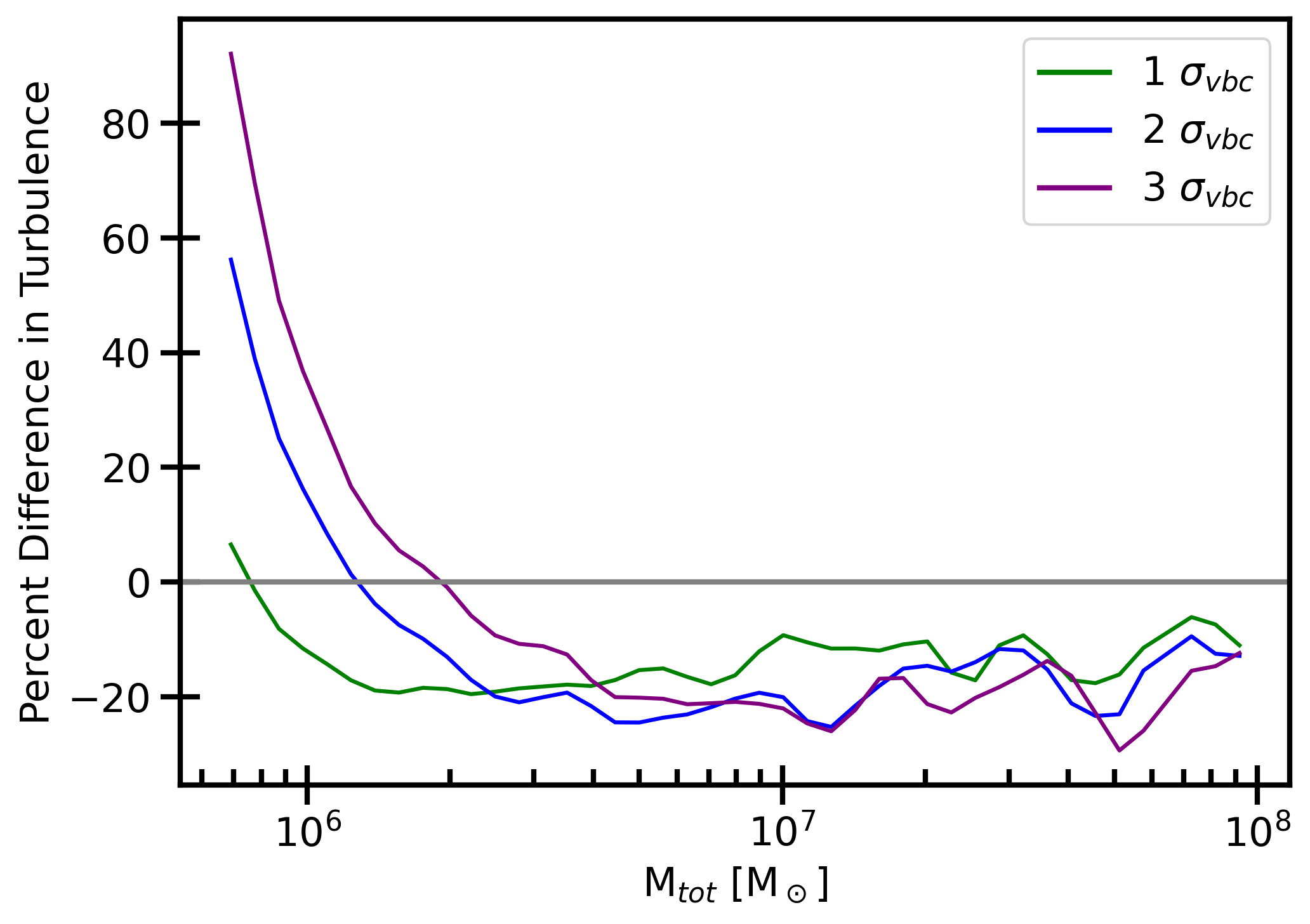}
    \caption{
    Percent difference in turbulent velocity between streaming ($v_{\rm bc} = 1\sigma_{\rm vbc}$, $v_{\rm bc} = 2\sigma_{\rm vbc}$, $v_{\rm bc} = 3\sigma_{\rm vbc}$) and no-streaming ($v_{\rm bc} = 0\sigma_{\rm vbc}$) simulations as a function of total halo mass. The percent difference is computed separately in each mass bin. The enhancement of turbulence grows with increasing streaming for low-mass halos ($M \lesssim 10^6\ \mathrm{M_\odot}$), while turbulence is suppressed at higher masses across all streaming strengths.
    }
    \label{fig:sigma}
\end{figure}

\autoref{fig:sigma} shows the percent difference as a function of total halo mass. The $v_{\rm bc} = 3\sigma_{\rm vbc}$ run exhibits both a stronger enhancement in turbulence in low-mass halos and a shift of the turnover mass to higher values. In contrast, the $v_{\rm bc} = 1\sigma_{\rm vbc}$ run shows a more modest enhancement, with the turnover beginning around $M \sim 7 \times 10^5\ \mathrm{M_\odot}$. For high-mass halos, we observe the expected inverse trend: turbulence is suppressed relative to the no-streaming case, and this suppression becomes weaker with decreasing $v_{\rm bc}$.

\end{document}